# Thermally induced generation and annihilation of magnetic chiral skyrmion bubbles and achiral bubbles in Mn-Ni-Ga Magnets

Bei Ding[1,2,3#], Junwei Zhang[2,4#], Hang Li[1,3], Senfu Zhang[2], Enke Liu[1,5], Guangheng Wu[1], Xixiang Zhang[2*] and Wenhong Wang[1,5*]

[1] Beijing National Laboratory for Condensed Matter Physics and Institute of Physics, Chinese Academy of Sciences, Beijing 100190, China

[2] Physical Science and Engineering, King Abdullah University of Science and Technology (KAUST), Thuwal 23955-6900, Saudi Arabia

[3] University of Chinese Academy of Sciences, Beijing 100049, China

[4] Key Laboratory for Magnetism and Magnetic Materials of Ministry of Education, Lanzhou University, Lanzhou 730000, People's Republic of China

[5] Songshan Lake Materials Laboratory, Dongguan, Guangdong 523808, China

[#] Contributions: Bei Ding and Junwei Zhang contributed equally to this work.

*Corresponding author.    E-mail: xixiang.zhang@kaust.edu.sa
wenhong.wang@iphy.ac.cn

**ABSTRACT:**

Magnetic chiral skyrmion bubbles and achiral bubbles are two independent magnetic domain structures, in which the former with equivalent winding number to skyrmions offers great promise as information carriers for further spintronic devices. Here, in this work, we experimentally investigate the generation and annihilation of magnetic chiral skyrmion bubbles and achiral bubbles in the Mn-Ni-Ga thin plate by using the Lorentz transmission electron microscopy (L-TEM). The two independent magnetic domain structures can be directly controlled after the field cooling manipulation by varying the titled angles of external magnetic fields. By imaging the magnetization reversal with increasing temperature, we found an extraordinary annihilation mode of magnetic chiral skyrmion bubbles and a non-linear frequency for the winding number reversal. Quantitative analysis of such dynamics was performed by using L-TEM to directly determine the barrier energy for the magnetization reversal of magnetic chiral skyrmion bubbles.

Nanometric magnetic textures have long been the subject of intensive study, continuing to be fueled by the promise of spintronic devices. A vortex-like topological spin texture, called skyrmion, is first theoretically predicted [1] and mostly observed [2-4] in magnets with non-centrosymmetric cubic crystal structure. Recently, the topological equivalent types of magnetic chiral skyrmion bubbles (SKBs) are found in centrosymmetric magnets with dipole-dipole interaction and uniaxial magnetic anisotropy [5-9]. Contrary to DMI-stabilized skyrmions with a fixed chirality, chiral SKBs in centrosymmetric magnets possess two degrees of freedom, i.e., chirality and vorticity [2]. The chirality is defined by the in-plane magnetization, which is rotated direction along the perimeter, while the vorticity is represented by a winding number $S$, which characterizes the topology of bubbles [10]. The winding number $S$ for magnetic bubbles in the centrosymmetric magnets varies with the defects of the domain wall, thus resulting in different magnetic topological state such as chiral biskyrmions ($S = 2$) [6, 11, 12], chiral SKBs ($S = 1$) [5, 13] and achiral bubbles (ABs) or hard bubbles ($S = 0$) [14, 15]. Since $S = 1$ SKBs are considered as skyrmions, the dynamics of metastable chiral SKBs including their generation and annihilation behavior as a function of external field and temperature, especially the relaxation time $\tau$, is an important issue for potential skyrmion-based applications.

The magnetic field and temperature are the most critical controllable external parameters for the skyrmions' generation and stability. For example, the dynamics of metastable skyrmions with a changing external magnetic field was studied indicating a unique aggregation and collapse between skyrmions and the conical phase.[16] Further experiments demonstrated that a hidden metastable skyrmion lattice could be obtained behind non-topological magnetic order by applying a sufficient high field cooling rate.[17] In addition, the recent study with the relaxation dynamics of zero field metastable skyrmions via thermal manipulation in the chiral magnet FeGe represented three distinct switching modes from skyrmions to stripes, suggesting a non-Arrhenius law behavior.[18] The creation of the skyrmion lattice with thermal assist has been studied in Pt/Co/Ta multilayer films.[19] However, the existing work is mostly restricted to skyrmions in chiral magnets, the dynamic evolution of the formation and annihilation of chiral SKBs remains elusive. Here, in this letter, we report on the generation of metastable chiral SKBs and ABs in the Mn-Ni-Ga thin plate via field cooling (FC) procedure. By tuning the titled angles of external magnetic fields, we investigated the dynamic annihilation of the two states with increasing temperature as a control parameter for the stability. Importantly, the chiral SKBs annihilation dynamics is directly observed by *in-situ* L-TEM representing an extraordinary evolution into the paramagnetic state.

The as-cast polycrystalline $(Mn_{1-x}Ni_x)_{65}Ga_{35}$ (x = 0.45) is the same as that used in our previous work.[6] We found that the formation of topological spin textures significantly depended on both the Mn-Ni-Ga crystal orientation and the applied magnetic field.[20] Further neutron scattering studies revealed the emergence of a non-collinear canted magnetic structure along the c-axis.[21] The Mn-Ni-Ga sample for the L-TEM observation was prepared by the mechanic polishing, and the magnetic domain evolution behavior with thermal manipulation was observed by using FEI Titan G2 60–300 equipped with a heating holder. The Curie temperature of the Mn-Ni-Ga thin plate is about 325 K, and the specific FC process is similar to our previous studies.[11,13] The specific FC manipulation was summarized as follows: first, the sample was heated up to 360 K, which was higher than Curie temperature $T_C \sim 325$ K; Second, a small perpendicular magnetic field was applied by increasing the objective lens current gradually in a very small increment; Third, the temperature was cooled down gradually from 360 K to 300 K; Finally, at 300 K, the perpendicular magnetic field was turned off. The experimental process was recorded by the charge coupled device (CCD) camera as shown in Supplementary **Movie S1 and Movie S2**. The external magnetic field was applied along the e-beam direction by controlling the current of the objective lens. To quantitatively analyze the domain structure, three images (under-, over- and in-focus) were acquired with a CCD camera and the in-plane magnetization distribution was reconstructed by the Qpt software based on the transport-of-intensity equation (TIE).

We first discuss the case at θ =0°; i.e., using the FC technique described, the magnetic field is applied normal to the (001) nano-sheet plane. The results are summarized in the form of a magnetic phase diagram in **Fig. 1(a)**. The spontaneous ground state of stripe domains remains unchanged when the magnetic field is lower than 300 Oe, shown in **Fig. 1(b)**. With the magnetic field increasing, the chiral SKBs appear. One can notice that the density of chiral SKBs firstly increase and then decrease with the increase of the magnetic field. The representative L-TEM image of chiral SKBs after an optimized 1000 Oe FC manipulation is shown in **Fig. 1(c)**. When a higher magnetic field of 2000 Oe is applied, the mixed stripes and chiral SKBs are shown up in **Fig. 1(d)**. This is because the magnetic field is too strong under this condition, the nucleation sites will be forced to agglomerate, thereby forming stripe domain again. As a whole, the magnetic field plays a critical role during the FC process, and an optimized magnetic field exists to generate the chiral SKB phase with the highest density.

We then investigate the formation of ABs under different tilted angles of magnetic fields during FC manipulation. The schematic experimental configuration is shown in **Fig. 2(a)** and **(c)**, in which the inclined magnetic field H is realized by tilting the nano-sheet, while the magnetic field is fixed along the direction of electron beam and the value is fixed at H = 1000 Oe. **Figure 2(b)** and **(d)** show the typical AB's arrangements at two typical tilted angles θ = 5° and 10°. We find that two kinds of bubbles could be transferred by tilting the sample. When the angle is smaller than 3°, the chiral SKBs form. With the increasing of the angle, most chiral SKBs evolve into ABs, shown in **Fig. 2(b)**. For θ > 10°, ABs are dominant (**Fig. 2(d)**). This effect is easily understood that the in-plane magnetic field component compels the magnetization orientated parallel to the in-plane magnetic field direction [22], which results in the transformation between chiral SKBs and ABs. Most interestingly, the hexagonal lattice persists up to θ ∼ 10°. The corresponding spin textures are displayed in **Fig. 2(e)**. The white arrows show the directions of the in plane magnetic inductions, while the black regions represent the domains with out-of-plane magnetic inductions. Chiral SKBs "1" and "2" with right- (C = +1) or left-handed chirality (C = -1) show up as black or white rings in the under-focused images both resulting in winding number $S = 1$, which is equal to that of skyrmions. ABs "3" composed of a pair of open Bloch lines, is characteristic of the domain structure commonly observed in ferromagnetic compounds with $S = 0$ (C = 0).

Based on these obtained chiral SKBs and ABs, the temperature dependent dynamic evolution of bubbles is investigated by using the *in-situ* L-TEM technique. The temperature is gradually increased from 301 K to 308 K, and the thermally activated excitation of bubbles winding number reversal ($S = 1$ to $S = 0$) has been observed (See Supplementary **Movie S3**). **Figure 3** are snapshots of raw L-TEM images for several temperature points with the exposure time 0.4 s per frame. As we increase the temperature, the majority of chiral SKBs dynamically and randomly reverse their chirality and transfer into ABs, while maintaining the hexagonal lattice (see **Fig. 3(a)-(g)**). With further increasing the temperature to near $T_C$, bubbles totally vanish into paramagnetic state. This phenomenon is consistent with previous study in the Ba-Fe-Sc-Mg-O [23]. To further investigate the temperature dependent dynamic evolution of chiral SKBs, 41 bubbles in the under-focus LTEM images were indexed as mentioned above. For each temperature, we calculated the averaged count of $S = 1$ and $S = 0$ bubbles based on the *in-situ* L-TEM video (see Supplementary **Movie S4**). **Figure 3(h)** shows the temperature dependence of the averaged statistic count of $S = 1$ and $S = 0$ bubbles. As the temperature increases, the count of chiral SKBs (S = 1) gradually decreases and sharply drops down to zero at 306 K. However, the ABs ($S = 0$) represent opposite behavior. For comparison, we perform the same procedure on the ABs which are chosen as the initial state (see Supplementary **Movie S5**), the corresponding thermal stability shown in **Fig. 3(i)-(l)**. Clearly, we can see that the count of $S = 1$ and $S = 0$ bubbles mostly maintain unchanged with the increase of the temperature. These analyses quantitatively demonstrate that the energy of chiral SKBs ($S = 1$) is lower than ABs ($S = 0$) at room temperature and thermally energy can active their transition. Noticeably, we find an extraordinary dynamic annihilation mode of chiral SKBs under thermally activation, that is, metastable chiral SKBs firstly transfer into ABs with the increase of the temperature and then totally collapse into paramagnetic state near $T_c$. This is very different from the collapse dynamics of skyrmions in chiral magnet FeGe[18].

To clearly understand the dynamic behavior of chiral SKBs and ABs, we focus on individual representative chiral SKB. We found that a chiral SKB permanently exhibits repeated reversal of the winding number at a given T, as shown in **Fig. 4(a)**. At a lower temperature, for example 300 K, the switching between the chiral SKB and the AB is infrequent and the chiral SKB state is more stable. In contrast, as the temperature increases up to 303 K, the winding number switches very quickly between 0 and 1. Upon increasing temperature to 305 K, the switching frequency slows down and the AB state is favorable. The switching process with the temperature further verified the annihilation mode of the metastable chiral SKBs. In essence, the behavior shows the relaxation dynamics. In **Fig. 4(b)**, we show the temperature dependence of the mean relaxation time $\tau_n$ for the chiral SKB and the AB, of which $\tau_n$ is calculated by the experiments shown in **Fig. 4(a)**.

To quantitatively evaluate the activation energy of a single chiral SKB, we use the Arrhenius law, $\tau = \tau_0 \exp\left(\frac{E_s}{k_B T}\right)$, where $\tau_0$ is the pre-exponential factor, $\kappa_B$ is the Boltzmann constant and $E_s$ is the activation energy. In estimating the order of magnitude of $E_s$, we assume a simple temperature-dependent activation energy $E_s/\kappa_B$ = a $(T_c - T)$ (See supplementary material for details).[17] From the fitting result (yellow line), we can estimate the thermal activation energy $E_s$ ($S = 1$) to be ~9.9x10$^{-20}$ J at 301 K. This value is in the same magnitude with the previously reported switching energy barrier skyrmions in Fe-Co-Si,[24] FeGe[16] and MnSi[17], but 2 order of magnitude lower than 10$^{-17}$ J of skyrmions in La-Ba-Mn-O[25]. As for the ABs, we find that the curve is poor fitted as a function of temperature using the Arrhenius equation, implying that activation energy of the AB represents a non-line variation with the temperature, which represents a similar behavior of metastable skyrmions in FeGe [13]. (See supplementary material for details). Based on the activation energy curve, the phenomenological free-energy landscapes of chiral SKBs and ABs can be sketched in **Fig. 4(c).** At room temperature, the energetically lower state is the chiral SKB. As the temperature increased, the activation energy of chiral SKBs appear linear decreasing while a subline decreases of ABs, thus resulting in the same energy of two kinds of bubbles. When the temperature is near $T_c$, ABs become more prominent with a lower energy.

In summary, the generation and annihilation of chiral SKBs and ABs with temperature have been clearly investigated in Mn-Ni-Ga by using L-TEM. The two residual magnetic states can be directly controlled after the FC manipulation by varying the angle of the magnetic fields. Systematic analysis of the *in-situ* L-TEM video demonstrates an extraordinary annihilation mode of chiral SKBs and a non-linear frequency for the winding number reversal with increasing temperature. The switching process are characterized by the temperature dependent activation energy showing the activation energy ~ $9.9 \times 10^{-20}$ J at 301 K. This work provides the basic information for controlling the topological magnetization texture and the stability.

See supplementary material for the calculation of thermal activation energy of chiral SKBs and ABs.

See supplementary movie for the evolution of magnetic domain structures as a function of temperature observed by using in-situ Lorentz-TEM.

This work is supported partially by the National Key R&D Program of China (Grant Nos. 2017YFA0303202 and 2017YFA0206303), the National Natural Science Foundation of China (Grant No. 11974406 and 51801087), the Key Research Program of the Chinese Academy of Sciences, KJZD-SW-M01 and the King Abdullah University of Science and Technology (KAUST) Office of Sponsored Research (OSR)

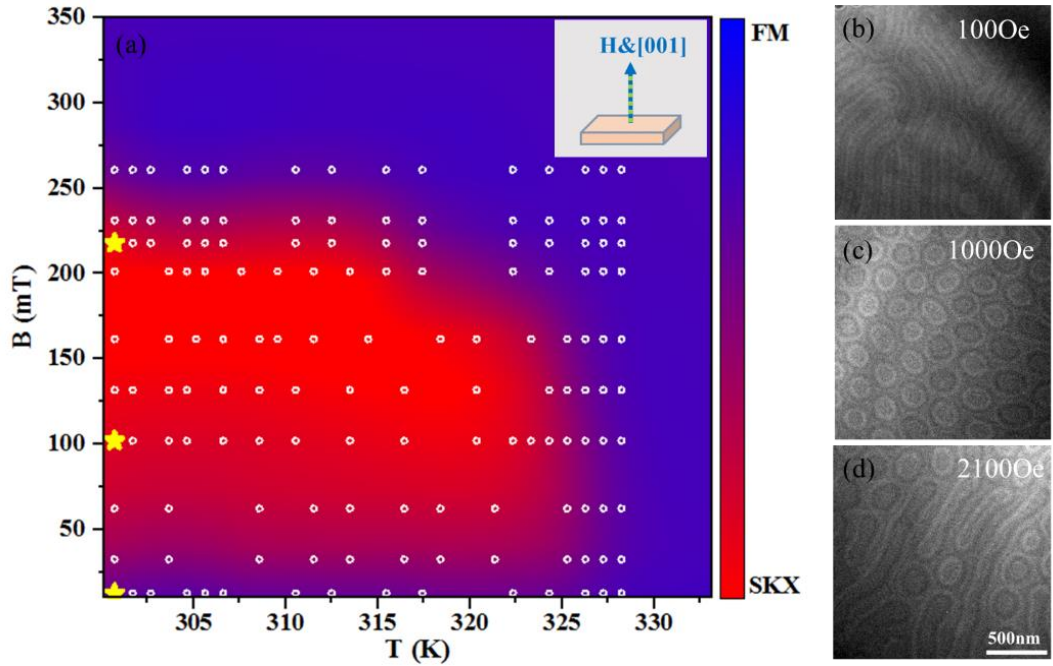

**Fig.1. Magnetic phase diagram of chiral skyrmion bubbles density with FC procedure in the magnetic field versus the temperature plane.** The magnetic field is applied normal to the (001) nano-sheet plane. Experimentally measured positions are marked by white open dots, and the representative L-TEM images of (b), (c) and (d) are labeled with star symbols. (b), (c) and (d) Stripes, chiral skyrmion bubbles and chiral skyrmion bubbles + stripes after FC procedure with magnetic field at 100 Oe, 1000 Oe and 2000 Oe. The scale bar is 500 nm.

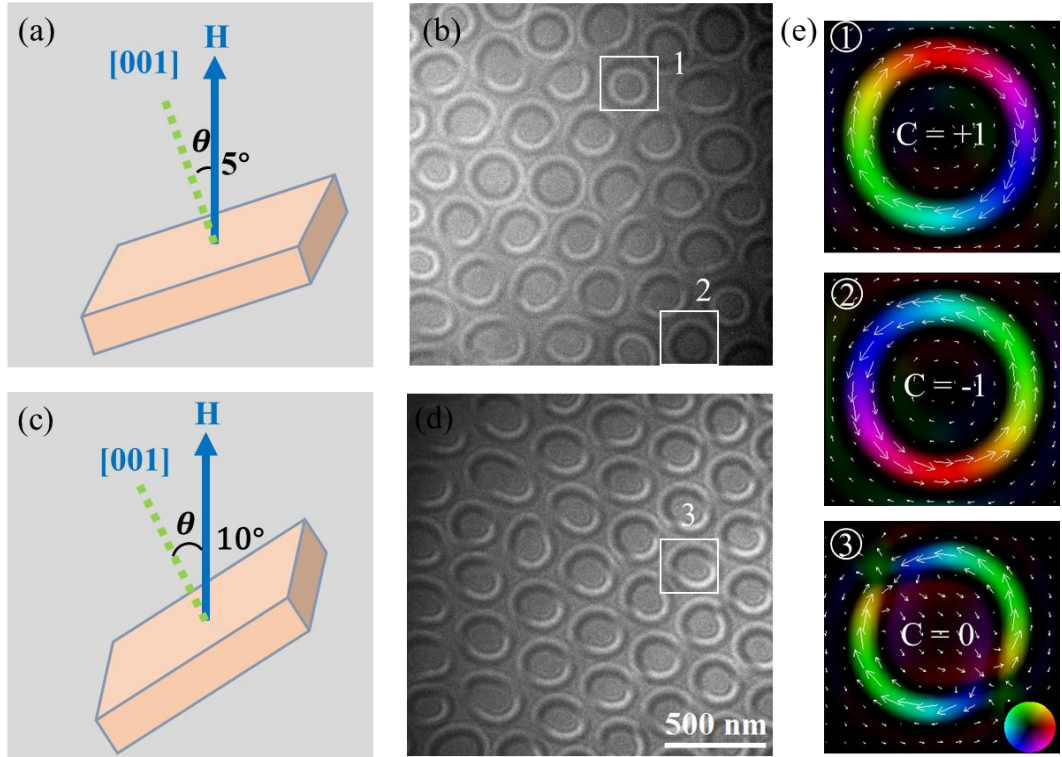

**Fig.2. Variation of topological magnetic state at room temperature under different tilted angles after FC manipulation.** (a) and (c) The schematically experimental configurations. θ is the tilted angle. (b) and (d) Under-focus L-TEM images of chiral skyrmion bubbles and achiral bubbles with tilted angles θ = 5° and θ = 10°. (e) In-plane magnetization distribution map obtained by TIE analysis for magnetic texture indicated in (b) and (d). Colors (the inset of panel (e) shows the color wheel) and white arrows represent the direction of in-plane magnetic induction, respectively. Panel 1 and 2 display the chiral skyrmion bubbles with C= +1 and C = −1 chirality both resulting in *S* = 1 and panel 3 shows the achiral bubbles with C = 0 (*S* = 0).

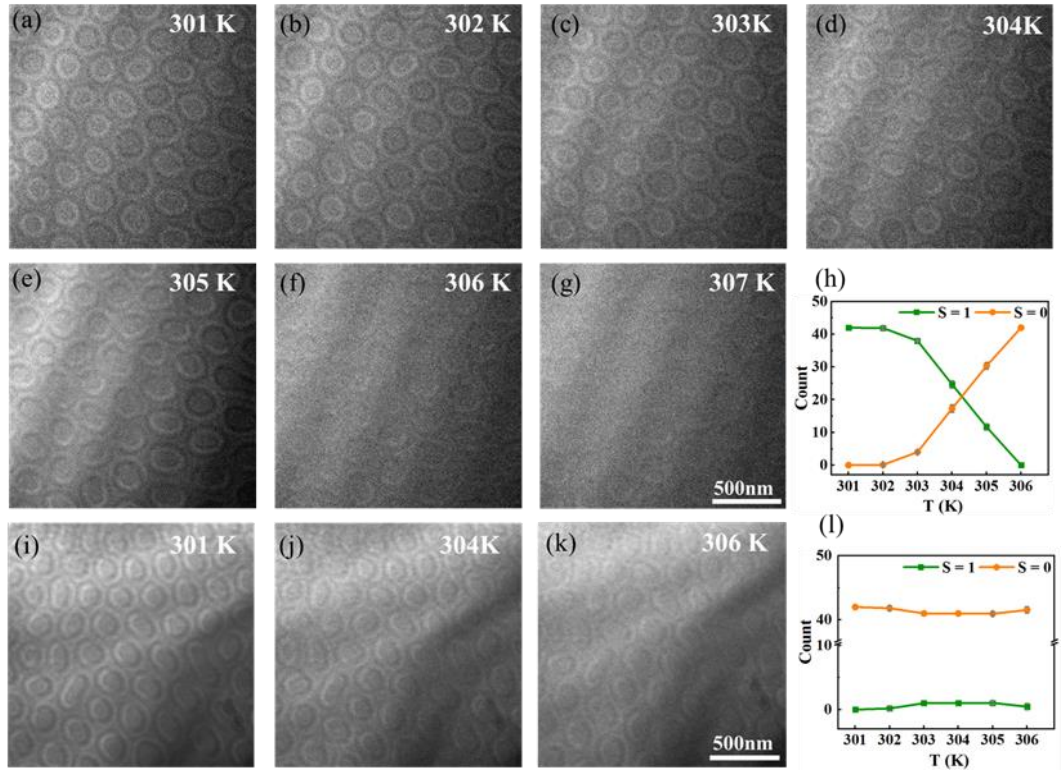

**Fig.3. Temperature dependence of zero field chiral skymrion bubbles and achiral bubbles.** (a) - (g) Chiral skyrmion bubbles are chosen as the initial state. Series of under-focus L-TEM images observed at zero magnetic field at 301 - 307 K. The chiral skyrmion bubbles with $S = 1$ show up as dark or white rings, while the achiral bubbles represent two arches ($S = 0$) in the L-TEM contrast images. Exposure time of each frame is 0.4 s. (h) Statistical averaged count of zero field chiral skyrmion bubbles and achiral bubbles with the increasing temperature. The error bar was indicated by the gray line. (i) - (k) Achiral bubbles are chosen as the initial state. Snapshots of L-TEM video observed at 301 K, 304 K and 306 K. (l) Statistical averaged count of zero field chiral skyrmion bubbles and achiral bubbles with the increasing temperature.

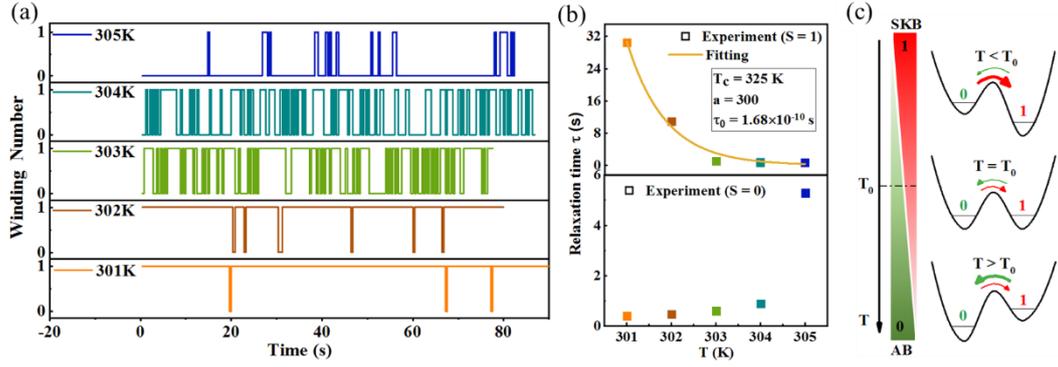

**Fig.4. Analysis of dynamic transition of a single bubble.** (a) Switching between $S = 1$ and $S = 0$ as a function of time in a specimen temperature range from 301 K to 305 K. (b) Temperature dependence of mean relaxation time $\tau_n$ for a single chiral skyrmion bubble and achiral bubble. The colorful squares represent the experimental statistic mean relaxation time $\tau_n$. The yellow line is fitted by using Arrhenius' law. (c) The schematic free-energy landscape of the temperature-dependent potential for chiral skyrmion bubbles ($S = 1$) and achiral bubbles ($S = 0$).